# Modulations of the Work Function and Morphology of a Single MoS$_2$ Nanotube by Charge Injection

Maja Remškar *[a], Janez Jelenc [a], Nikolai Czepurny [b], Matjaž Malok [a], Luka Pirker [a,c], Rupert Schreiner [b], Andreas K. Hüttel [d]

a. Solid State Physics Department, Jozef Stefan Institute, Jamova ulica 39, SI-1000 Ljubljana, Slovenia; *Corresponding author, E-mail: maja.remskar@ijs.si

b. Faculty of General Sciences and Microsystems Technology, OTH Regensburg, 93053 Regensburg, Germany

c. Department of Electrochemical Materials, J. Heyrovsky Institute of Physical Chemistry, Dolejskova 3, Prague 8, 182 23 Czech Republic

d. Institute for Experimental and Applied Physics, University of Regensburg, 93053 Regensburg, Germany

**A current was injected into a single MoS$_2$ nanotube using an atomic-force-microscopy probe. The trapped electrons and holes caused a partial collapse of the nanotube and its helical twisting. The topography changes can be explained by the reverse piezoelectric effect, the theory of which was proposed for chiral nanostructures. Work-function modifications were observed, which were dependent on the polarity of the injected carriers.**

MoS$_2$ belongs to the group of transition-metal dichalcogenide crystals with a layered structure. It has recently attracted a lot of attention as a quasi-two-dimensional graphene analogue. Besides a plate-like morphology, spherical and cylindrical shapes have been reported.[1-5] The unique growth of MoS$_2$, directly from the vapour phase, leads to MoS$_2$ nanotubes (NTs) with lengths up to several millimetres, and diameters from a few tens of nanometres up to several micrometres.[2] Due to their slow synthesis, which runs for several weeks, growth takes place close to the chemical equilibrium, leading to a high degree of structural perfection. These NTs support the appearance of whispering-gallery modes,[6] and represent a quantum-confinement environment for single-electron conductance.[7] The crystal structure of MoS$_2$ consists of S-Mo-S molecular layers with covalent bonds between intra-layered atoms, and weak Van der Waals interactions among these molecular layers. Due to the absence of dangling bonds on the basal (001) surfaces, an inertness and the absence of adsorbents are expected. In reality, the exposed (001) planes of MoS$_2$ flat crystals were found to be n-doped due to the spontaneous formation of sulphur vacancies, causing surface-electron accumulation, which was nearly four orders of magnitude higher than in the inner bulk.[8] The electrical conductance in MoS$_2$ is highly anisotropic and varies from sample to sample, with 0.16 S/cm to 5.12 S/cm along the basal planes and 1.02·10$^{-4}$ S/cm to 5.89·10$^{-4}$ S/cm in the direction perpendicular to the (001) planes.[9] The ratio between the conductance along the planes and that perpendicular to them is at least 10$^3$. In thin flakes (<10 μm), the conductance was found to be much larger and strongly dependent on the thickness,[8] i.e., it increased from 11 S/cm to 360 S/cm with a decrease in the thickness from 385 to 33 nm. In comparison to freshly cleaved MoS$_2$ flakes without the surface states caused by desulfurization,[8] surface studies of MoS$_2$ NTs, which cannot be cleaved without structural destruction, are challenging. Surface integrity and cleanliness at the atomic level are assumed, but they cannot be proven for each tube under investigation.

As opposed to bulk 2H-MoS$_2$, which is not piezoelectric due to its centrosymmetric structure with an anti-parallel stacking sequence, thin flakes of MoS$_2$ composed of an odd number of S-Mo-S molecular layers show a piezoelectric response in line with that theoretically predicted[10] and experimentally observed.[11,12] A piezoelectric output was observed for flakes consisting of one, three and five monolayers, although its intensity decreased with an increase in the number of layers. A large value of the piezoelectricity was observed in 3R-MoS$_2$ and two piezoelectric coefficients $d_{33}$ and $d_{13}$ were determined as ≈0.9 and ≈1.6 pm/V, respectively. These coefficients were found not to be dependent on the thickness of the flakes.[13]

Piezoelectricity was also theoretically predicted for a tubular geometry of single-layer BN due to the alternating group-III and group-V elements, which lower the symmetry, and by a process of elastic coplanar deformation produce a non-zero polarization.[14] A longitudinal piezoelectric response was proposed to exist for the zigzag NTs, but only under extension or compression along a NT axis, while in the armchair NTs the dipole moment was coupled only to the torsional strain. In chiral tubes, the elastic energy contains the product of the axial and torsion strains. Therefore, a tensile stress induces torsion and vice versa, and a cross-section of such a chiral tube should become deformed with respect to the ideal cylindrical geometry along its length as a result of a long-range elastic force. In $WS_2$ NTs, an in-plane sliding ferroelectricity was reported recently.[15] It was explained by the piezoelectric nature of the asymmetric structure of the $WS_2$ monolayers. Due to the change in piezoelectricity in the layers of different chirality, the shear force exceeding the friction force enables a mutual sliding and subsequent deformation of the NT. A programmable photovoltaic effect was achieved using a prior bias, which alters the rectification behaviour of the $WS_2$ NT devices.

This study presents the first experimental observation of the shape and work-function (ϕ) modulations caused by a charge injection into a single $MoS_2$ NT. Atomic force microscopy (AFM) was used to inject the current and to observe the topography. Kelvin probe microscopy (KPFM) was used to record the changes in the contact potential difference (CPD) between the AFM tip and the NT before and after the injections of current. The shape modifications of the NT can be explained with the radial and torsional components of the reverse piezoelectric effect.

The $MoS_2$ NTs under investigation and $MoS_2$ substrate were synthesized in a chemical transport reaction at 1010 K inside a quartz ampoule using iodine as the transport agent.[2] Such NTs always grow in chiral mode, with the chiral angle in the range 10° to 16°, and most frequently around 16°.[3] A homogeneous diameter of the tube along a relatively long length (up to several tens of microns) shows that the lattice structure does not only consist of wrapped $MoS_2$ monolayers, but that the layers are self-terminated. It is the elimination of the edge dangling bonds that stabilizes the cylindrical shape of these NTs. The proper stacking of the molecular layers depends on the diameters of the tubes. The rhombohedral 3R stacking has been observed in tubes with diameters of more than 200 nm, while the hexagonal 2H stacking is typical for narrower tubes.[4] The NT selected for the experiment, with a diameter of around 50 nm, belongs to the group of narrow nanotubes that crystallize in the 2H stacking. A detailed HRTEM structural characterization was performed on another NT of the same diameter, i.e., 50 nm (Fig. 1a). The wall thickness was about 30 % of its diameter. The corresponding diffraction pattern taken from the whole diameter of the NT is an overlap of electrons diffracted from the basal planes, oriented from parallel to perpendicular to the electron beam. The chiral angle was 13° (Fig. 1b). The diffraction pattern can be indexed according to 2H-$MoS_2$ (ICSD 03-065-0160). The wall consisted of 26 molecular layers of the same chirality (Fig. 1b, c). Traces of amorphous material are visible on the outer surface of the NT. These result from the TEM sample-preparation process that uses drop-casting of a dispersion of the NTs in ethanol.

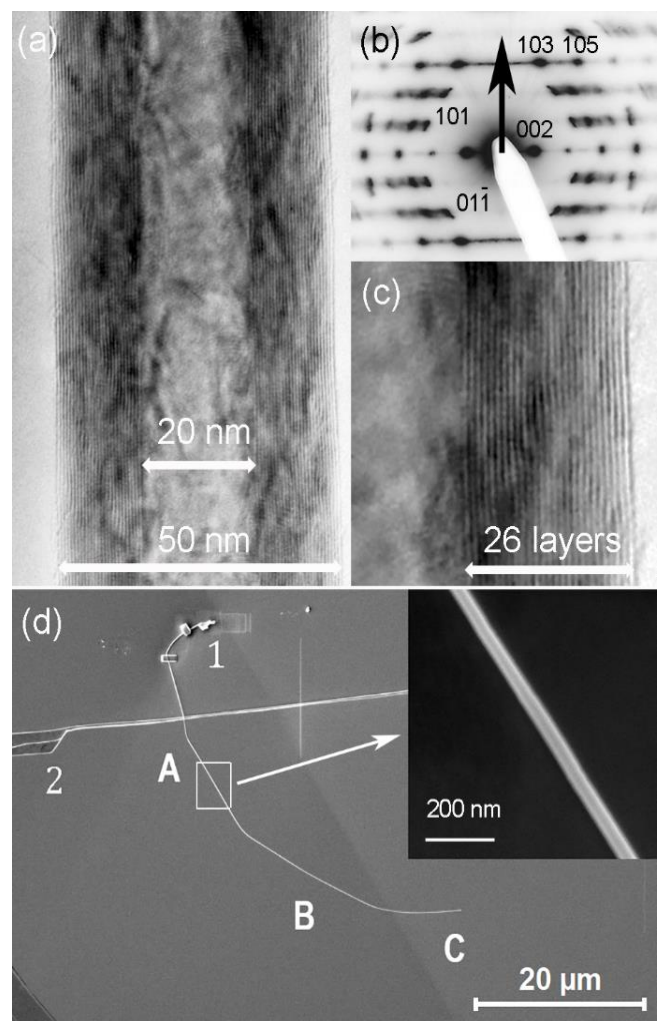

Figure 1: a) HRTEM image of a $MoS_2$ NT, around 50 nm in diameter; b) corresponding diffraction pattern revealing 2H stacking; arrow points to the NT's axis; c) wall consisting of 26 molecular layers; d) SEM image of the $MoS_2$ NT on a single-crystal $MoS_2$ substrate, with diameter 53 nm near the Pt contacts (A) and 45 nm in the central part (B). A, B, C: positions of the topography and CPD measurements; B, C: positions of the charge injections; 1: Pt contacts fixating the nanotube; 2: $MoS_2$ substrate surface steps. Arrow points to enlarged image of the NT.

The NT selected for the charge injection was deposited on a $MoS_2$ single crystal using an OmniProbe manipulator 200 (with a sharp tungsten tip) in a FEI Helios Nanolab 650 SEM-FIB microscope ($10^{-6}$ mbar). Platinum contacts were deposited to keep the NT in place (Fig. 1d).

The charge injections were performed inside an ultra-high-vacuum atomic force microscope (VT-AFM Omicron, Germany). A Kelvin-probe force microscopy (KPFM) module was used to measure the contact potential difference (CPD) between the AFM tip, the NT, and the $MoS_2$ single-crystal substrate. Both, $MoS_2$ NT and $MoS_2$ substrate can contain traces of iodine used in the transport reaction, therefore only relative values of the CPDs were taken into account. The current was injected in the following way. The NT was first positioned using the non-contact AFM (nc-AFM). The Pt/Ir-coated silicon AFM tip (Type NSG10-Pt, NT-MDT, Spectrum Instruments) with a tip-curvature radius of around 35 nm was used. By switching to the scanning tunnelling microscopy (STM) mode, the forced oscillations of the cantilever were stopped and the tip was auto-approaching towards the sample until the pre-defined tunnelling current (200 pA) was established. Charge carriers were injected for 2 min under a bias from +2 V to +8 V and from -2 V to -8 V in 2 V steps. The current of 200 pA was controlled by the feedback loop. After the injections of holes at positive voltages, a break of 48 h was taken and then the electrons were injected at negative biases. Breaks other than 48 hours were also used in individual experiments. Before and after each series of injections, the shape and CPD of the NT were recorded at three places (Fig.1): near the Pt contacts (position A), in the central part (position B), and at the very end of the NT (position C). The changes in the CPD profiles were normalized in such a way that the CPD of the substrate was zero. Any tilting of AFM topography images was corrected using the max flatness tilt in SPIP 6.7.5.

A schematic of the changes to the work function of the NT ($\phi_{MoS2}$) as a function of the external bias during the current injections is shown in Fig. 2. Due to the larger $\phi$ of the tip ($\phi_{tip}$) with respect to the $\phi_{MoS2}$ (a), electrons were transferred from the NT to the AFM tip under the tunnelling contact (b). The injection of electrons (c) caused a decrease of $\phi MoS_2$ with respect to the zero bias (d), while the injection of holes increased $\phi MoS_2$ (e).

**Injections at the position C and observation of their effects at positions B and A**

The first series of injections was performed at the very end of the NT, which was only attached to the substrate with the Van der Waals (VdW) interaction (position C). First, holes were injected at biases ranging from 1 V to 10 V. After 24 h, electrons were injected at -2V, -4V, -6V, and -8V. The effects of these injections were studied as a function of the distance from the injection spot, i.e., around 39 μm away (position A) and around 18 μm (position B). The topography profiles (Fig. 3a) reveal that the shape of the NT was modified along the entire length between A and C. At A (red), the height of the profile peak increased by around 1 nm (3.8 %), while its full width at half maximum (FHWM) decreased by 56 nm (28 %). At B (not shown), the height of the profile peak increased by around 2 nm (10 %), while its FHWM decreased by 30 nm (15 %). Narrowing of the profiles indicated that the NT was deformed from its ideal, cylindrical shape, but it did not move away from the substrate. Due to long range electric forces one can assume that the shape

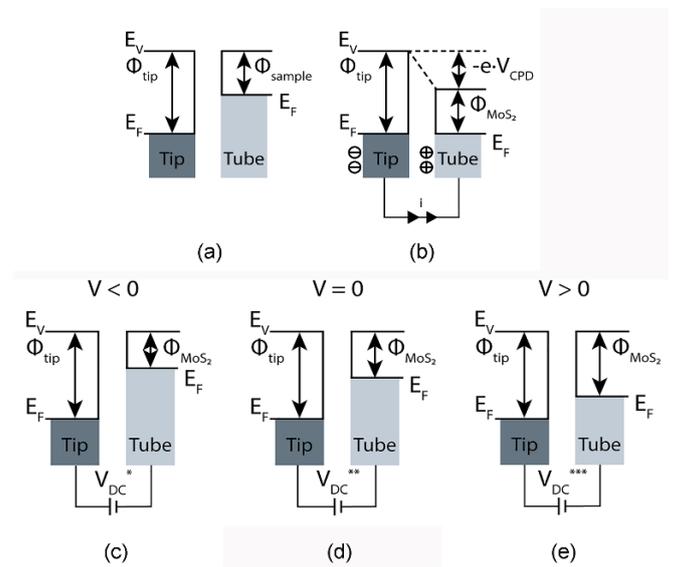

Figure 2: Electronic energy levels of the AFM tip (Tip) and of $MoS_2$ NT (Tube) for the cases: before injections tip and NT are separated (a) or in electric contact (b), and after injections at external negative bias (V<0) is applied to the tip (c); No bias/no injection (V=0) (d); injection at external positive bias (V>0) (e). $E_v$ is the vacuum energy level; EF is Fermi-energy level; $\phi_{tip}$ and $\phi_{MoS2}$ are work functions of the tip and the NT, respectively.

deformation occurs also at the interface with the substrate.

The deformation stress was not the same along the NT as was resolved from the Kelvin profiles (Fig. 3b). Before the injections, the CPD profile at A was parabolic with a minimum value that was 38 mV (black) and 96 mV (at B-not shown) lower than on the substrate. This difference between positions A and B could be interpreted as an unintentional contamination of position A with Pt, which has a relatively large work function (5.6 eV to 6.1 eV).[16] Carbon-based contamination during the sample preparation inside SEM cannot be excluded. After injections, the shapes of the CPD at positions A and B were changed. In the CPD line profiles, a peak appeared in the central part (up to 50 mV in height), which was surrounded by two minima of different depths. At A (red), the left-hand minimum is deeper (-100 mV), while at B (blue), the right-hand minimum is deeper (-150 mV). This difference, together with the different degrees of asymmetry between the left- and right-hand minima is explained by the chiral deformation of the NT.

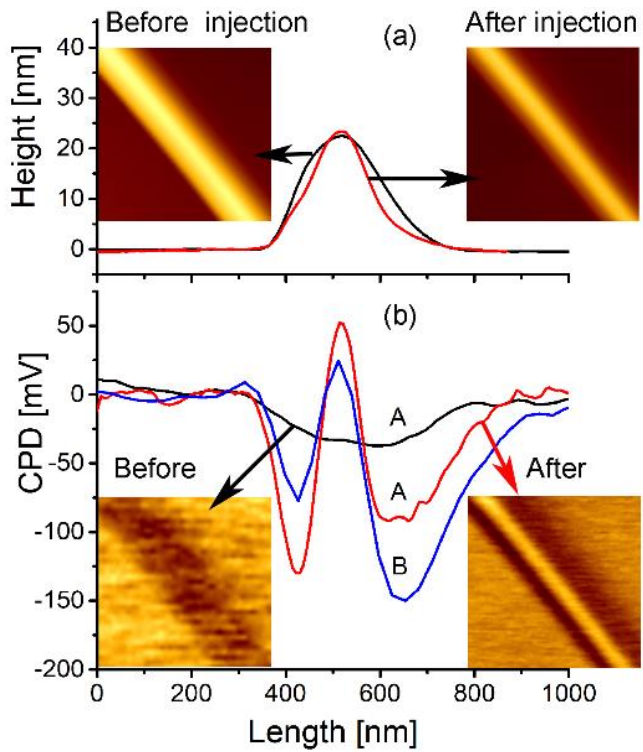

Figure 3: Changes of the MoS$_2$ NT at positions A and B caused by injections of the charge carriers at position C: a) Topography and (b) Kelvin microscopy images with corresponding line profiles before charge injection (black) and after the last electron injection at -8 V bias: A (red), B (blue). Size of the images: 1 μm × 1 μm. Line profiles taken along the scan lines.

## Injections at the position B and observation of their effects at position B

Before the first injection, a typical "holes-trapped" shape of CPD was recorded (Fig. 4b; black) with a central peak of 150 mV and two side minima. This shape remained from the previous injections at position C. It revealed that some holes are trapped in the NT or at its interface with the substrate, showing a kind of charge-memory effect. The absence of charge recombination is explained by the build-up of a local p-n junction at the substrate–NT interface due to different work functions. The MoS$_2$ substrate represents the p-side of this junction, while the NT represents the n-side. This "holes-trapped" state was then defined as the initial state for the next injection series.

Initially, holes were injected. The shape of the NT, which was already modified from the cylindrical geometry by the previous injections at position C, deformed to a shape with a distinguished central peak and two shoulders (Fig. 4a; red). The central peak in the line profile decreased by 20 %, while the FHWM of the whole profile increased by 25 %. The injection of holes compressed the NT in the direction towards the substrate and widened it. In addition, the compression caused a wrinkling of the profile as a result of the simultaneous elongation and compression of the sub-cylinders with different diameters. Different elastic constants in bulk MoS$_2$, i.e., $c_{11}$ along the basal planes (23.8 GPa) and $c_{33}$ perpendicular to them (5.2 GPa)[17] resist the isotropic reverse piezoelectric compression of the cylindrical shape. In addition, the chiral structure of the lattice makes the deformation of the NT even more complex. For any further observation of the shape deformations, it is important to note that the left-hand shoulder was higher than the right-hand one. The shape of the AFM tip was not changed during the injection test. The line profile of the topography of the NT near the surface step of the MoS$_2$ substrate (Fig. 1d) was kept parabolic.

After a 48-h break, electrons were injected at -2 V, -4 V, -6 V, and -8 V. The shape of the NT was changed in a way that the left-hand shoulder decreased and the right-hand one increased (Fig. 4a-blue). This is a clear indication that the NT was exposed to a torsional strain, which turned the NT in an anti-clockwise direction (looking from the top). The angle of rotation was 33°±0.5°, as estimated from the positions of the side peaks with respect to the centre of the NT at half the height of the central peak (Fig. 4a). After 48 h, holes were injected again. The already-compressed NT continued to rotate in the same direction as during the injection of electrons. The final injection of holes at +8 V caused the NT's shape to change drastically and the right-hand shoulder dominated the profile. The central peak decreased by around 70 %. The NT became heavily deformed into a semi-ribbon shape. The corresponding nc-AFM images are shown in the insets of Fig. 4a.

The CPD profiles show two qualitatively different, but typical, shapes. After the injection of holes, a peak in the central part of the NT dominates the CPD profile ("holes-trapped" shape), while after the injection of electrons, a deep valley in the CPD profiles is visible in the central part ("electrons-trapped" shape).

The first injection of holes caused the suppression of the central peak in the CPD profile by 40 % (Fig. 4b; red). The side valleys became shallower but wider. The CPD in the valleys was 150 mV (left-hand valley) and 190 mV (right-hand valley) below the CPD of the substrate. A typical hole-trapped shape appeared. After the injection of electrons (Fig. 4b; blue), a deep minimum of 225 mV occurred, surrounded by peaks of around 70 mV. The injection of holes was then repeated (Fig. 4b; pink). The central peak in the CPD again increased, but the ratio between right- and left-hand minima did not recover to the situation after the first injection of holes (red). This relation reflects the degree of irreversibility of the topography changes caused by the torsional component of the reverse piezoelectricity.

With the aim to explore the reversibility of the rotation and the shape-memory effect, two consecutive injections of electrons at -8 V were made at 1-h intervals, and 24 h later, holes were injected at +8 V (Fig. 5). The first injection of electrons caused the NT to turn in the clockwise direction (blue arrow). The left-hand shoulder in the profile increased with the voltage, but even after the final injection at -8 V it remained smaller than the right-hand one (light blue). The second injection of electrons at -8 V (blue) rotated the NT further in the clockwise direction. The three-peaks profile of the NT reoccurred. It is obvious that these two electron pulses did not completely turn the NT back to its initial orientation. Then the injection of holes followed at +8 V. The NT abruptly turned in anti-clockwise direction (red arrow) back to the orientation

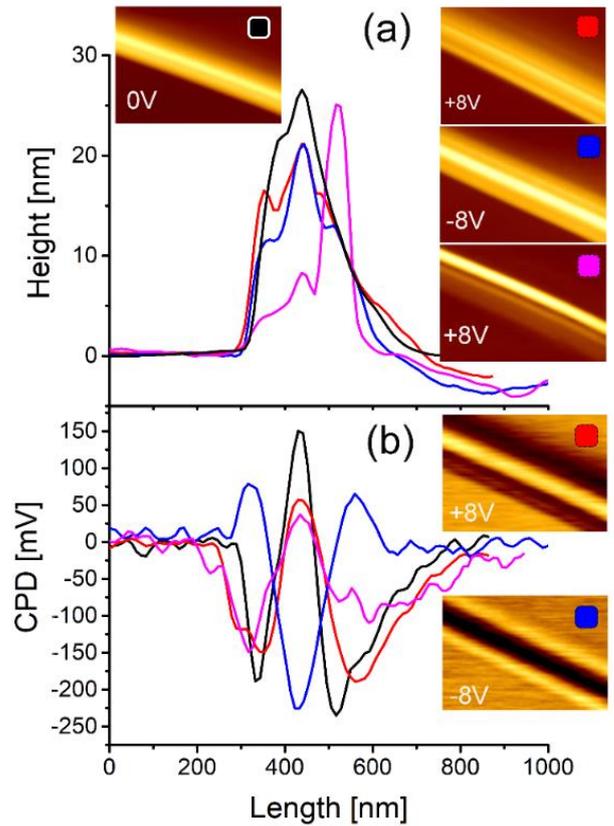

Figure 4: Changes to the MoS$_2$ NT at position B: a) Topography and b) Kelvin images with some corresponding profiles: initial state (black); after the first injection of holes at +8 V (red); after injection of electrons at -8 V (blue); after second injection of holes at +8 V (pink). Length of the images: 1 µm. Line profiles taken along the scan lines.

(red) that was similar to that before the second injection of electrons.

This rotation of the NT in a direction that alternates from clockwise due to the injection of electrons to anti-clockwise due to the injection of holes, indicates the torsional component of the reverse piezoelectric effect. At the voltages used in our experiment, the torsional deformation was not completely reversible. In addition, the NT was also progressively collapsing. The fact that the shape could not be recovered to the cylindrical geometry leads to the conclusion that besides elastic deformations, which were stabilized by the charge retention, plastic deformations also occurred through the creation of structural defects.

The CPD profiles revealed a charge-memory effect, which could be concluded from the following observations. After the first injection of electrons at -8 V (Fig. 5b-light blue), the typical shape of the holes-trapped profile with the central peak surrounded by two valleys was preserved. The central peak even increased by 45 % with respect to the height after the last injection of holes at +8 V (Fig. 4b-pink), while the right-hand valley deepened. This failure of the perfect transition from the holes-trapped to the electrons-trapped shape could indicate a kind of potential barrier for the transport of electrons due to the trapped holes. The second electron injection performed 1 h later was needed to finally cause a typical electrons-trapped feature (Fig. 5b-blue) with a deep minimum (200 mV) in the central part of the NT, surrounded by one peak (61 mV) on the left-hand side and two small peaks on the right-hand side (-5 mV, -55 mV). The following injection of holes at +8 V (red curve) immediately restored the typical holes-trapped shape of the CPD. The right-hand minimum in the CPD (-88 mV) became shallower than after the last injection of holes (-187 mV), while the central peak (190 mV) was the largest among all the CPD profiles in these experiments. The fact that two electron pulses were needed to annihilate the trapped holes and only one pulse of holes was sufficient to restore the holes-trapped shape might be explained by the fact that the holes are the main charge carriers in natural molybdenite ($MoS_2$).[9]

The occurrence of inverse piezoelectricity is explained by the charge retention at the NT–substrate interface due to the different work functions of the $MoS_2$ substrate and the $MoS_2$ NTs. The junction between the NT and the substrate forms a local p-n hetero-junction of two semiconductors. The injections strongly influenced the work function of the NT through enforced charge retention and the creation of an internal electric field. The injection of electrons made the NT more n-type, while the injection of holes made the NT less n-type or even p-type. An inhomogeneous work-function distribution along the NT diameter reveals an internal charge separation that is stabilized with the p-n junction at the NT–substrate interface. Therefore, no part of the NT is charge neutral. Different values of the work function for both sidewalls are explained by the torsional deformation, which changes the relative orientation of the molecular layers with respect to the substrate's normal and affects the internal strain. The injection of electrons intensified the p-n junction present before the injection in a way that more electrons were transferred from the NT to the $MoS_2$ substrate and the Fermi levels were equalized. Since $MoS_2$ is a good hole conductor, the electrons in the substrate gradually recombined with the holes from the grounded substrate. This makes electron retention less permanent compared to hole retention.

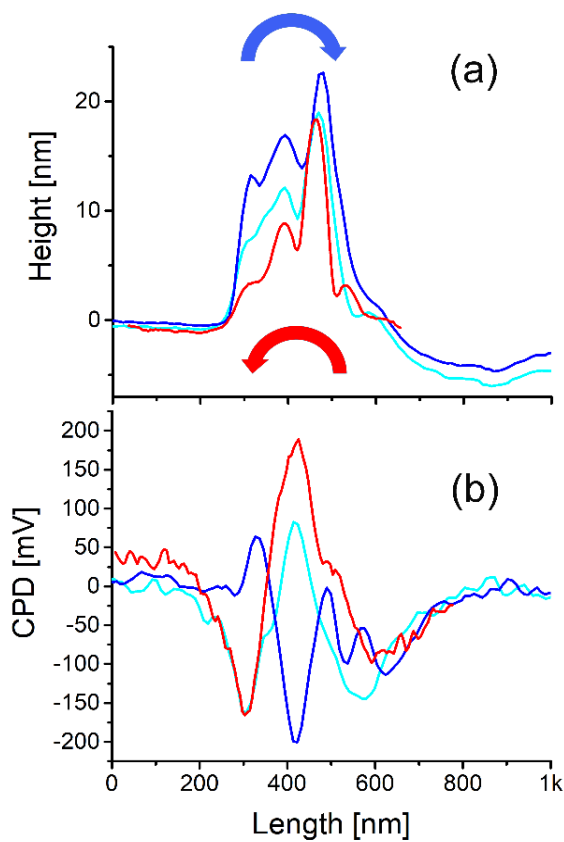

Figure 6: Rotation of the $MoS_2$ NT due to the injection of charge at position B: a) topography and b) Kelvin profiles after the first (light blue) and the second (blue) injections of electrons at -8 V, and after the injection of holes at +8 V (red). Curved arrows mark the direction of NT rotation due to the injection of electrons (blue) and holes (red).

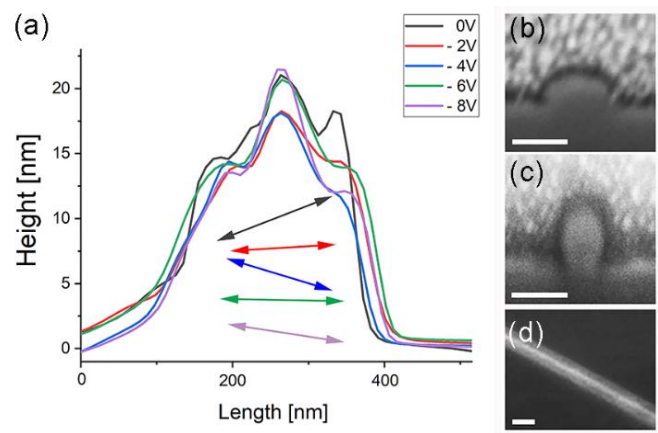

Figure 5: a) AFM topography profiles taken at position B after injections of electrons at -2 V (red), -4 V (blue), -6 V (green), and -8 V (violet) in comparison with the initial state (black); SEM micrographs of the NT after completed current injections: (b, c) cross-sections, and (d) the top-view. Scale bars: 50 nm (b, c); 100 nm (d).

The complexity of the torsional deformation is shown in Fig. 6a, which presents some profiles taken at the same place on the NT after consecutive electron injections at -2, -4, -6, and -8 V. The profiles are aligned using the middle peaks, while the positions and relative heights are represented by arrows. Because of the long-range elastic forces, the deformation at a certain place depends on the strain and plastic deformations at other parts of the NT. Nevertheless, rotation in the clockwise direction is visible, as are changes to the NT's width due to twisting. The trapped charges causing internal electric fields and deformation due to reverse piezoelectric effect, do depend on the diameter of particular molecular layers.[15] Therefore, the particular molecular layers inside the NT's wall with different circumference and energy of curvature were exposed to different degree of deformation. Both, the radial and torsion components of the stress were different in adjacent layers. The interaction among the layers weakened already during the first current injection and the initial cylindrical shape became wrinkled forming the three-peaks profile (Fig. 4a). With the aim to confirm the torsional deformation of the NT, it was cut by the FIB at different points along its length after completed current injection experiments. The SEM micrographs of two cross-sections (Fig. 6b, c) clearly demonstrate deviations from cylindrical symmetry and confirm the torsional deformation of the NT. In addition, the top view (Fig. 6c) reveals an increased concentration of secondary electrons along the NT's axis, which agrees with the model of wrinkling of the NT's wall, where thicker band of material contributes to higher density of secondary electrons.

In conclusion, electrons and holes were injected into a chiral $MoS_2$ nanotube. The captured charge caused a deformation of the NT from cylindrical to a semi-elliptical shape. The NT became helically twisted along its axis and progressively plastically deformed. These shape changes are explained by a coupling of the axial and torsional reverse piezoelectricity with a strength that depends on the diameter of particular molecular layers, the chirality of the tubular structure and the history of previous injections of charge carriers. The work function of the NT's sidewalls decreased after the injections of holes and increased after the injections of electrons. In contrast, the work function of the NT in its central part increased with the injection of holes and decreased with the injection of electrons. A charge-memory effect was observed for the trapped holes. These results are important for use of $MoS_2$ NTs in quantum physics, for the development of transducers or torsional resonators, and for understanding piezoelectricity and its reversible effect in chiral nanotubes.

**Author Contributions**: M.R. (Funding acquisition, Supervision-lead, Writing-original draft); J.J. (Methodology-lead, Investigation-lead); N.C. (Investigation, Formal analysis); M.M. (Conceptualization, Methodology-minor); L.P. (Methodology, Visualization-minor); R.S. (Supervision), A.K. H. (Writing-review & editing).

## Conflicts of interest

There are no conflicts to declare.

## Notes and references

Acknowledgements: The authors thank the Slovenian Research Agency for financial support via grants P1-0099 and PR-11224, and Deutsche Forschungsgemeinschaft for financial support via grants Hu 1808/4-1 (project id 438638106) and Hu 1808/6-1 (project id 438640730).